\documentclass[12pt]{iopart}
\usepackage{iopams}

\newcommand{\be}{\begin{equation}}
\newcommand{\ee}{\end{equation}}
\newcommand{\ben}{\begin{eqnarray}}
\newcommand{\een}{\end{eqnarray}}
\newcommand{\ra}{\rangle}
\newcommand{\la}{\langle}

\usepackage{iopams,epsf}  
\begin{document}

\title[Influence of anisotropy on the 
zero-temperature phase transition in the  $J-J'$ model
]{Influence of Ising-anisotropy on the 
zero-temperature phase transition in the square lattice 
spin-$\frac{1}{2}$ $J-J'$ model
}
\author{R. Darradi\dag,
J. Richter\dag and S.E. Kr\"uger\ddag}  

\address{

\dag Institut f\"ur Theoretische Physik, Universit\"at Magdeburg,
      P.O. Box 4120, D-39016 Magdeburg, Germany

\ddag IESK, Kognitive Systeme, Universit\"at Magdeburg,
      P.O. Box 4120, D-39016 Magdeburg, Germany
}

\begin{abstract}
We use a variational mean-field like approach, 
the coupled cluster method (CCM) and exact diagonalization
 to investigate the ground-state order-disorder transition 
for the square  
lattice  spin-half XXZ model with two different nearest-neighbor 
couplings $J$
and $J'$. Increasing $J' > J$ the model shows   in the isotropic Heisenberg limit  
a second-order
transition from  semi-classical N\'eel order to a quantum paramagnetic
phase with enhanced local dimer correlations on the $J'$ bonds at about
$J'_{c}
\sim 2.5 \ldots 3 J$. 
This transition is driven by the quantum competition between $J'$ and $J$.  
Increasing the anisotropy parameter $\Delta> 1$ we diminish the quantum
fluctuations and thus the degree of competition. 
As a result the transition point $J'_{c}$ is shifted to  larger values.
We find indications for a linear increase of $J'_{c}$ with $\Delta $, 
i.e. the transition disappears in the Ising limit 
$\Delta \to \infty$.

\end{abstract}

\section{Introduction}

The study of zero-temperature phase transitions driven by quantum fluctuations 
has been a subject of great interest to
physicists during the last decade, see \cite{sachdev,qpt_ri01,uzun03,ri04} 
and references therein. 
For order-disorder quantum phase 
transitions we basically need the interplay bet\-ween the interparticle
interactions and fluctuations. A canonical model to study quantum
phase transitions is the spin-half Heisenberg
model with competing interactions in two dimensions. 
Competition  between bonds appears in frustrated systems.
 Besides frustration,
 there is another mechanism  weakening the 
ground-state N\'eel order in 
 Heisenberg antiferromagnets,
 namely the competition of non-equivalent nearest-neighbor (NN) 
bonds leading to the
 formation of local singlets of two (or even four) 
coupled spins.
By contrast to frustration, which yields
competition in quantum as well as in classical systems, this type of
competition is present only in quantum systems. 

Recent experiments on  
$\mathrm{SrCu}_2(\mathrm{BO}_3)_2$
\cite{fat,has} and on $\mathrm{CaV}_4\mathrm{O}_9$\cite{rac,azi} 
demonstrate the
existence of gapped
quantum paramagnetic ground states in (quasi-)two-dimensional 
Heisenberg systems and have stimulated various theoretical studies of
quantum spin lattices with competing interactions.
A famous example for competition in  a frustrated Heisenberg antiferromagnet
is the spin-half
$J_1-J_2$ model on
the square lattice, where the frustrating $J_2$ bonds plus quantum
fluctuations lead to a second-order transition from N\'eel ordering to a
disordered quantum spin liquid, see e.g. \cite{doucot88,abd,foa,bishop98,moh}. 
On the other hand,
it has been predicted that frustration may lead
to  a first-order transition in
quantum spin systems in contrast to a second-order transition in the
corresponding classical model \cite{lar,ila,mos,krueger00}.
An example for competition without frustration is the 'melting'
of semi-classical N\'eel order by local singlet formation in  
Heisenberg systems with two non-equivalent nearest-neighbor bonds
like the bilayer antiferromagnet \cite{sandvik94,gros95}, 
the $J-J'$ antiferromagnet 
on the square lattice \cite{krueger00,sing88,krueger01} and 
on the depleted square
 (CaVO) lattice \cite{troyer96,troyer97}.

In the above mentioned 
papers\cite{krueger00,sandvik94,gros95,sing88,troyer96,troyer97} the strength 
of quantum fluctuations is tuned by variation of the exchange bonds.
Alternatively, the strength of quantum fluctuations can be tuned by 
the anisotropy
$\Delta$ in an XXZ model.  

In this paper we study the influence of the Ising anisotropy 
on the zero-temperature magnetic order-disorder transition 
for the $J-J'$ spin-half XXZ antiferromagnet 
on the square lattice. 
We use a variational 
mean-field like approach (MFA), the coupled cluster method (CCM) and exact
diagonalization (ED). We mention that the CCM, being 
one of the most
powerful methods of quantum many-body theory, has previously been
applied to various  quantum spin systems with much success
\cite{bishop98,krueger00,krueger01,rog_her90,bishop94,re,zeng98,bishop00}.

\section{Model}
\label{model}
\begin{figure}
\begin{center}\epsfxsize=20pc
\epsfbox{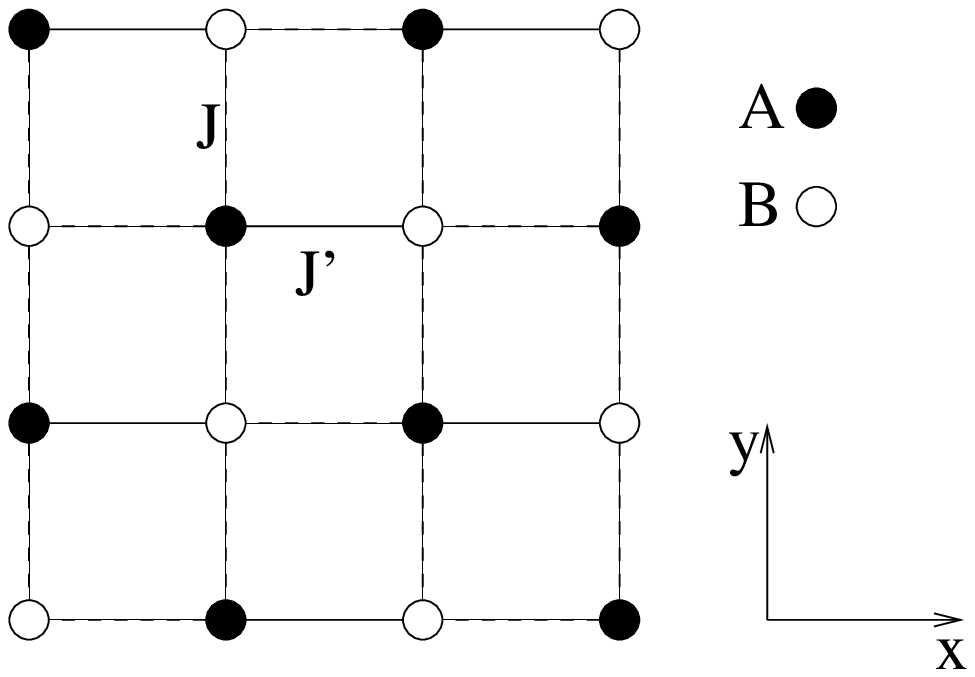}
\end{center}
\caption{\label{fig1} 
Illustration of arrangement of bonds in the 
$J-J'$ model on the  square lattice
(Eq.(\ref{eq1})): 
$J$ -- dashed lines;   $J'$ -- solid lines; A and B
characterize the two sublattices of the classical N\'eel ground state.
}
\end{figure}

We consider an anisotropic spin-$\frac{1}{2}$ Heisenberg (XXZ) model on the 
square lattice
with two kinds of nearest-neighbor bonds $J$ and $J'$, as shown in
Fig.\ref{fig1}:
\begin{eqnarray}
\label{eq1}
H = J\sum_{<ij>_1} \left ( s_i^{x}s_j^{x}+
s_i^{y}s_j^{y}+\Delta s_i^{z}s_j^{z} \right ) 
+ J'\sum_{<ij>_2}\left (s_i^{x}s_j^{x} + s_i^{y}s_j^{y} +
\Delta s_i^{z}s_j^{z} \right ) .
\end{eqnarray}
The sums over $<ij>_1$ and $<ij>_2$ run over 
the two kinds of nearest-neighbor bonds, respectively (cf. Fig.\ref{fig1}).
Each square-lattice plaquette consists of three $J$ bonds and one $J'$
bond. We consider antiferromagnetic bonds $J'\ge J>0$, i.e. there is no
frustration in the model. In what follows we set $J = 1$ and consider the
Ising anisotropy $\Delta \ge 1$ and $J'$ as the parameters of the
model.
Since there is no frustration the classical ground state is the
two-sublattice N\'eel state. 

\section{Methods}
\subsection{Variational mean-field approach (MFA)}

For the square-lattice antiferromagnet ($J=J'$) the ground state is N\'eel
ordered. The corresponding  uncorrelated mean-field state 
is the N\'eel state
$|\phi_{MF_1}\rangle  = |\hspace{-0.13cm} \uparrow \rangle
|\hspace{-0.13cm} \downarrow \rangle 
|\hspace{-0.13cm}\uparrow \rangle|\hspace{-0.13cm}\downarrow \rangle 
\ldots$ . In the limit $J' \to \infty$ and for finite $\Delta$ the ground
state approaches a rotationally invariant 
product state of local pair singlets (valence-bond state) 
$|\phi_{MF_2}\rangle  = 
\prod_{i \in A}  \left\{|\uparrow_{i} \rangle|\downarrow_{i+\hat{x}} 
\rangle
-|\downarrow_{i} \rangle|\uparrow_{i+\hat{x}} \rangle \right\} /\sqrt{2} \; ,\quad$ 
where 
$i$ and $i+\hat{x}$ correspond to those sites which cover the $J'$ bonds. 
In order to describe the transition between both states, we consider 
an uncorrelated
product state  of the form\cite{krueger00,gros95}
\begin{equation}
\label{eq2}
|\Psi_{var}\rangle=\prod_{i\in A} \frac{1}{\sqrt{1+t^2}}[ |\uparrow_i
\downarrow_{i+\hat{x}}\rangle - t|\downarrow_i \uparrow_{i+\hat{x}}\rangle ].
\end{equation}
The trial function $|\Psi_{var}\rangle$ 
depends on the variational parameter $t$ and 
interpolates between the valence-bond state $\; |\phi_{MF_2}\rangle\; $ 
realized for
$t=1$ and the N\'eel state $\; |\phi_{MF_1}\rangle\; $ realized 
for $t=0$. By minimizing $E_{var}=\langle\Psi_{var}|H|\Psi_{var}\rangle$
with respect to the variational parameter $t$  we obtain
\begin{eqnarray}
\label{eq3}
\frac{E_{var}}{N} = \left\{ \begin{array}{ll}
-\frac{1}{24\Delta} \left( J'^2+3\Delta^2J'
+9\Delta^2 \right ) &\; \textrm{ for $J'\leq
3\Delta$}\\
-\frac{1}{8}J'(\Delta+2) & \; \textrm{ for $J'>3 \Delta$ }.
\end{array} \right.  
\end{eqnarray}  
The relevant order parameter describing the N\'eel order is the sublattice
magnetization
\begin{eqnarray}
\label{eq4}
M_s = \langle \Psi_{var}| s^z_{i \in A}|\Psi_{var}\rangle
= \left\{ \begin{array}{ll}
\frac{1}{2}\sqrt{1-(J'/3 \Delta)^2}  &\; \textrm{ for $J'\leq
3\Delta$}\\
0  &\; \textrm{ for $J' > 3\Delta .$}
\end{array} \right. 
\end{eqnarray}
$M_s$ vanishes at the critical value   $J'_c=3\Delta J$. The
corresponding critical index is the mean-field index
$1/2$. Eq. (\ref{eq3}) may be rewritten in terms of $M_s$ as 
\be
E_{var}/N =
-\frac{1}{8}J' \Delta - \frac{1}{4}J' \sqrt{1-4M_{s}^2} - \frac{3}{2}  \Delta
M_s^2.
\ee 
We expand $E_{var}$ up to the fourth order in $M_s$ near the
critical point and we find a Landau-type expression, given by 
\be 
E_{var}/N =
-\frac{1}{8}J' (\Delta+2) +  \frac{1}{2}(J' - 3 \Delta)M_s^2  +\frac{1}{2} J'
M_s^4.
\ee     

\subsection{Coupled cluster method (CCM)}

The CCM formalism is now briefly considered,  for
further details the interested reader is referred to Refs.
\cite{zeng98,bishop00}. 
The starting point for the calculation of the many-body ground state is
a normalized reference or model state $|\Phi\rangle$. 
For our model the appropriate
reference state is the N\'eel state.
To treat each site equivalently we perform a rotation of the local axis of
the up spins such that all spins in the reference state align in the same
direction, namely along the negative $z$ axis.
After this transformation we have 
$\; \; |\Phi\rangle  = |\downarrow \rangle|\downarrow \rangle 
|\downarrow \rangle|\downarrow \rangle
\ldots \; \;$. 
Now we define a set of multi-spin creation operators $C_I^+=s_r^+ \; , \;
s_r^+ s_l^+ \; , \; s_r^+ s_l^+ s_m^+\; , \; \ldots \;$ . 
The choice of the $C_I^+$ ensures that $\langle \Phi| C_I^+ = 0 =
C_I|\phi\rangle$, where $C_I$ is the Hermitian adjoint of $C_I^+$. 
The CCM parametrizations of the ket and bra ground 
states are then given by
\be\label{ket} 
  |\Psi\ra =e^S|\Phi\ra, \quad S=\sum_{I\neq 0}{\cal S}_IC_I^+ ,\ee
\be\label{bra} 
  \la\tilde\Psi|=\la\Phi|\tilde Se^{-S}, \quad \tilde S=
1+\sum_{I\neq 0}\tilde {\cal S}_IC_I .\ee
The correlation operators $S$ and $\tilde S$ contain 
the  correlation coefficients
${\cal S}_I$ and $\tilde {\cal S}_I$ which have to be calculated.
Using the Schr\"odinger equation, $H|\Psi\ra=E|\Psi\ra$, we can now write 
the ground-state energy as $ 
    E=\la\Phi|e^{-S}He^S|\Phi\ra$ . 
The order parameter 
is calculated by $
 M_s = -\la\tilde\Psi|s_i^z|\Psi\ra$ .

To find the  correlation coefficients 
${\cal S}_I$ and $\tilde {\cal S}_I$ we have
to require that the expectation value $\bar H=\la\tilde\Psi|H|\Psi\ra$
is a minimum with respect to 
${\cal S}_I$ and $\tilde {\cal S}_I$.
This formalism is exact if we take into account  all possible multispin 
configurations in
the correlation operators $S$ and $\tilde S$. 
Of course, this is impossible for our model.
Thus we have to use approximation schemes to truncate the expansion
of $S$ and $\tilde S$ in the Eqs.~(\ref{ket}) and (\ref{bra}). The most common scheme
is the LSUB$n$ scheme, where we include only $n$ or fewer correlated spins
in all configurations (or lattice animals in the language of graph theory)
which span a range of no more than $n$ adjacent 
lattice sites. 

To improve the results it is useful to extrapolate the 'raw' CCM-LSUB$n$ 
results to the limit $n \to \infty$.
Although no exact scaling theory for results of LSUB$n$ approximations is
available, 
there are empirical indications\cite{krueger00,bishop94,zeng98,bishop00}
of scaling laws for  the order parameter for antiferromagnetic 
spin models. 
In accordance with those findings we use 
$ M_s(n)=M_s(\infty)+a_1(1/n)+a_2(1/n)^2$ to extrapolate to $n \to \infty$.
Vanishing $M_s(\infty)$ determines the critical point $J'_c$.
The values for $J'_c$  obtained by extrapolation of the $LSUBn$ results
for $M_s$ are,
however, found to be slightly too large \cite{krueger00}. 
We may also consider the inflection
points of the $M_s(J')$ curve for the $LSUBn$ approximation, assuming that the
true $M_s(J')$ curve will have a negative curvature up to the critical point. 
We might expect
that (for increasing $n$) the inflection point $J'_{inf}$ 
approaches the critical point $J'_c$.
Thus determining  the 
inflection points for the $LSUBn$ approximation
again we can extrapolate to the limit $n\to\infty$ 
using a scaling law $ J'_{inf}(n)=J'_{inf}(\infty)+b_1(1/n)+b_2(1/n)^2$
and interpret $J'_{inf}(\infty)$ as the critical value $J'_c$.
\subsection{Exact diagonalization (ED)}
In addition to the variational mean-field approach and the CCM we use ED 
to calculate the order parameter. We consider finite
square lattices of $N=8,10,16,18,20,26,32$ sites and employ periodic
boundary conditions. The relevant order parameter for finite systems is
the square of the sublattice magnetization $M_s^2$, here defined as 
$M_s^2 =
\langle [ \frac{1}{N}\sum^N_{i=1}\tau_i{\bf s}_i ]^2 \rangle 
$ with the staggered factor $\tau_{i \in A} =+1$, $\tau_{i \in B} =-1 $.
For the finite-size 
scaling of $M_s^2$
we use the standard three-parameter 
formula\cite{neuberger89,hasenfratz93,oitmaa94,sandvik97}
$M_s^2(N) =  M_s^2(\infty)+ c_1N^{-1/2} + c_2N^{-1}$.
The critical value $J'_c$ is that point where
$M_s^2(\infty)$ vanishes.

\section{Results}
\begin{figure}
\begin{center}\epsfxsize=30pc
\epsfbox{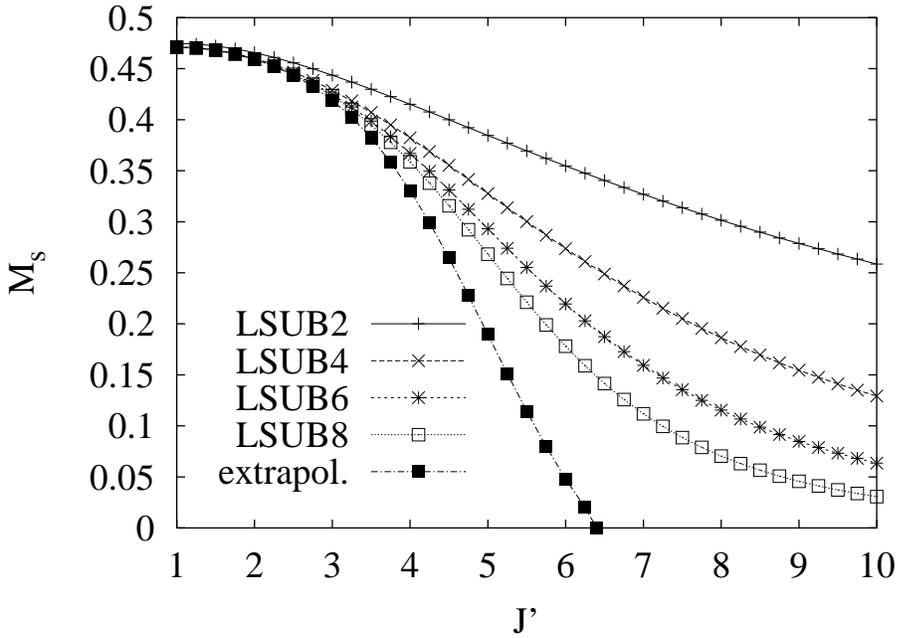}
\end{center}
\caption{\label{fig2} Sublattice magnetization $M_s$ versus $J'$ for 
$\Delta=2$  
using coupled cluster method  (CCM), see text.}
\end{figure}

\begin{figure}
\begin{center}\epsfxsize=30pc
\epsfbox{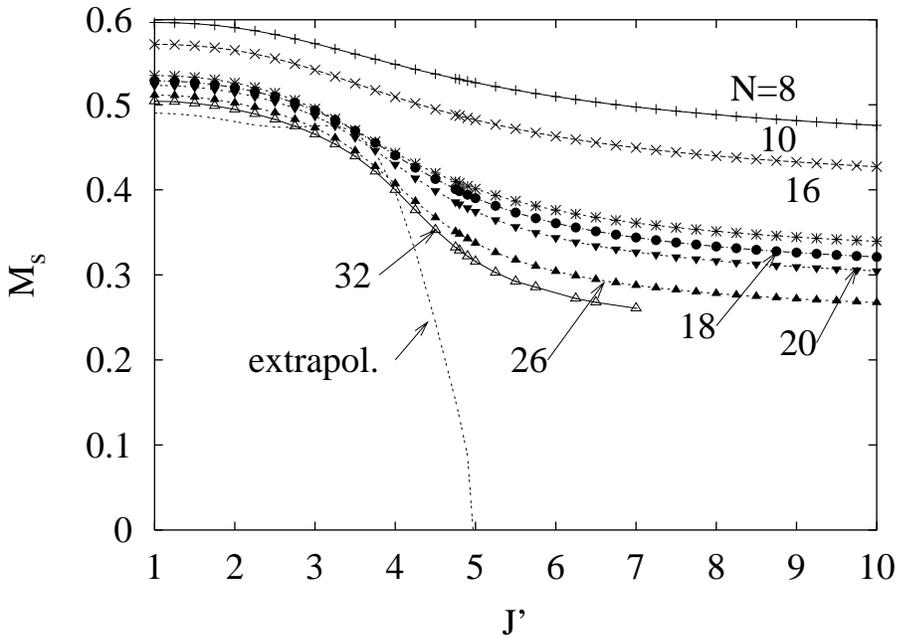}
\end{center}
\caption{\label{fig3} Order parameter versus $J'$
for $\Delta=2$ using exact diagonalization of finite lattices of different
sizes $N$, see text.}
\end{figure}

To illustrate the behavior of the oder parameter in dependence on
$J'$  we present  $M_s(J')$ calculated by CCM (Fig.\ref{fig2}) and by
ED (Fig.\ref{fig3}) and the resulting extrapolated values for 
a particular value of Ising anisotropy $\Delta=2$.
Notice that corresponding results for the isotropic Heisenberg case ($\Delta=1$) can
be found in Ref.\cite{krueger00}.

From the extrapolated order parameters one gets the 
critical values for $\Delta=2$: $J'_c=6.46$ (CCM) and $J'_c=4.97$ (ED). The
mean-field value is $J'_c=6$. The inflection points of the $M_s(J')$
curves in Fig.\ref{fig2} are 
$J'_{inf}(n)= 5.57 $  (LSUB2), $ 5.42 $ (LSUB4),   $5.26$  (LSUB6), $5.11$
 (LSUB8) leading to an extrapolated value  of $J'_c= J'_{inf}(\infty)=
 4.66$.
As mentioned above, the extrapolation of the CCM results of the 
order parameter
tends to overestimate the critical value and yields the largest $J'_c$. 
This is connected with the change
of the curvature in the $M_s$-$J'$ curve in the vicinity of the
critical point, cf. Fig. \ref{fig2}. 
Therefore the critical value $J'_c$ taken form the inflection points
seems to be more realistic.     
Obviously, the difference in $J'_{inf}$ between the LSUB$n$ approximations
is small and the extrapolated value is quite close to 
the value for
LSUB8. This statement holds for all values of $\Delta$. E.g. for $\Delta=4$
one finds $J'_{inf}= 10.37 $  (LSUB2), $ 10.29 $ (LSUB4),   
$10.13$  (LSUB6), $9.94$
 (LSUB8) leading to an extrapolated value  of  $J'_c= J'_{inf}(\infty)=
 9.41$.

\begin{figure}
\begin{center}\epsfxsize=30pc
\epsfbox{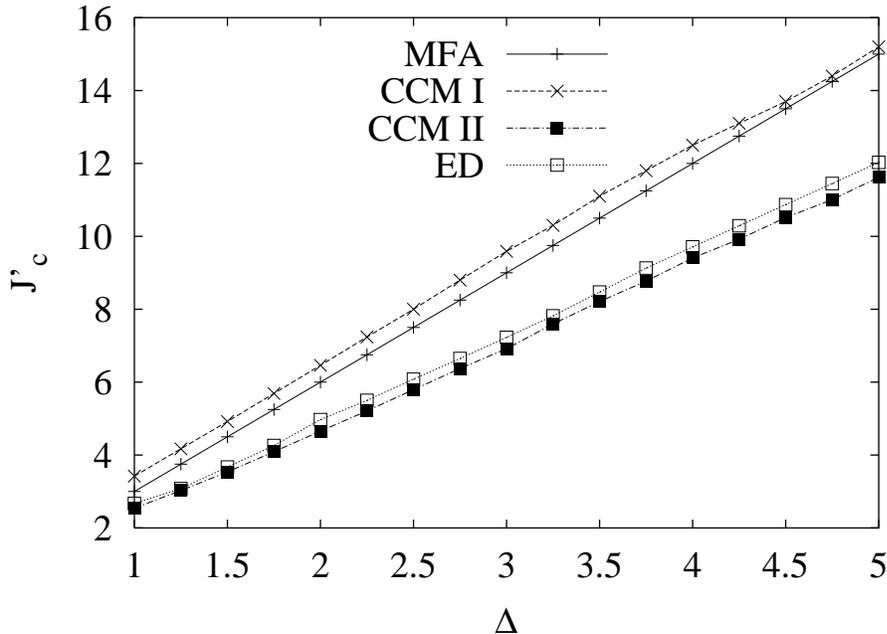}
\end{center}
\caption{\label{fig4} The critical points $J'_c$ versus anisotropy parameter
$\Delta$ using mean-field approach (MFA), CCM with extrapolation of the order
parameter (CCM I), CCM with extrapolation of the inflection points
(CCM II) and exact diagonalization (ED).
}
\end{figure}

Our results for the  critical point $J'_c(\Delta)$ obtained by MFA, CCM and
ED are collected in Fig.\ref{fig4}. 
We find that
the CCM results obtained by the extrapolation of the order parameter  are in good 
agreement with the MFA data. On the other hand,  there is an excellent
agreement between the CCM results obtained by the extrapolation of the 
inflection points and
the ED results obtained by the extrapolation of the order parameter.
Clearly we see indications for a linear increase in $J'_c$ as predicted by
mean-field theory. 

We mention that the curves shown in Fig.\ref{fig4} cannot be extrapolated to
$\Delta < 1$. Similar to the effect of the Ising anisotropy ($\Delta > 1$)
one rather expects an increase of $J'_c$ due to XY anisotropy, i.e. for 
$0 \le \Delta  < 1$. Indeed for the pure XY $J-J'$ model ($\Delta=0$) 
  the critical value 
was estimated to  $J'_c=4.56J$\cite{tomczak01a}.

\section{Summary}

We have studied the zero-temperature magnetic ordering in a square-lattice
spin-half anisotropic Heisenberg (XXZ) model with two kinds of
nearest-neighbor exchange bonds $J$ and $J'$, see Fig.\ref{fig1}. 
In particular we discuss the influence of the Ising anisotropy $\Delta$
on the
position of the quantum critical 
point $J'_{c}$ separating the phase with 
semi-classical N\'eel order ($J' < J'_{c}$) and the quantum
paramagnetic phase without magnetic long-range order ($J' > J'_{c}$). 
For this we 
calculate the order parameter within a variational mean-field approach, 
the coupled cluster method  and exact
diagonalization of finite lattices up to $N=32$ sites. 
We find in good approximation a linear  relation 
$J'_{c}(\Delta) \propto \alpha \Delta $ ($\Delta \ge 1$) with 
$\alpha \sim   
2.3 \ldots 3.0 $. This result can be attributed to the reduction  of
quantum spin fluctuations with increasing Ising anisotropy. In the
pure Ising limit ($\Delta \to \infty$) the only 
remaining $z-z$ terms in the
Hamiltonian (\ref{eq1}) commute with each other, i.e. no quantum spin 
fluctuations
are present and, consequently, 
the critical point disappears in the pure Ising limit. 
 \\

\ack
R. Darradi thanks the Land Sachsen-Anhalt for financial support.
The authors are indebted to J. Schulenburg for assistance in numerical
calculatuons.

\end{document}